\documentclass[aps,twocolumn,preprintnumbers]{revtex4}
\usepackage{amssymb}

\usepackage{amsmath}
\usepackage{graphicx}
\usepackage{dcolumn}
\usepackage{bm}

\begin{document}

\title{Quantum information processing based on $^{31}$P nuclear spin qubits\\
in a quasi-one-dimensional $^{28}$Si nanowire}
\author{Issai Shlimak$^{1}$ and Israel Vagner$^{2,3,}$\footnote{deceased}}
\affiliation{$^{1}$Minerva Center and Jack and Pearl Resnick Institute of Advanced
Technology,\\
Department of Physics, Bar-Ilan University, Ramat-Gan 52900, Israel\\
$^2$Research Center for Quantum Communication Engineering,\\
Department of Communication Engineering, Holon Academic Institute of
Technology, Holon 58102, Israel\\
$^{3}$Grenoble High Magnetic Fields Laboratory, Max-Planck-Institut f\H{u}r
Festk\H{o}rperforshung and CNRS, F-38042 Grenoble Cedex 9, France}

\begin{abstract}
We suggest a new method of quantum information processing based on the
precise placing of $^{31}$P atoms in a quasi-one-dimensional $^{28}$Si
nanowire using isotope engineering and neutron-transmutation doping of the
grown structures. In our structure, interqubit entanglement is based on the
indirect interaction of $^{31}$P nuclear spins with electrons localized in a
nanowire. This allows one to control the coupling between distant qubits and
between qubits separated by non-qubit neighboring nodes. The suggested
method enables one to fabricate structures using present-day
nanolithography. Numerical estimates show the feasibility of the proposed
device and method of operation.
\end{abstract}

\maketitle

\section{Introduction}

There is great scientific and commercial interest in the development of
quantum computation (QC) and the creation of computational devices based on
the principles of quantum logic. Several different schemes for QC have been
proposed to date (see, for example, Ref.~\onlinecite{DiVincenzo} and
references therein). One of the exciting avenues, potentially compatible
with the vast fabrication capabilities of modern semiconductor technology,
relies on the encoding of information in the electron or nuclear spins
present in semiconductor nanostructures, leading to a spin-based
semiconductor quantum computer \cite{Loss,PVK98,Kane98,Kane1,Vrijen}.

The most developed model of the nuclear spin quantum computer is the Kane
suggestion \cite{Kane98} to use a precisely located array of phosphorous
donors introduced into Si. In this proposal, the nuclear spin 
${}^{1}\!/{}_{2}$ of $^{31}$P is used as a qubit, while a donor electron
together with an overlying gate (A-gate) separated from the donor by a 
SiO$_{2}$ or Si$_{0.85}$Ge$_{0.15}$ barrier, provides single-qubit operation
using an external magnetic field and pulses of radio-frequency radiation.
The interqubit coupling is determined by the overlap of the electron wave
functions and is controlled by metallic gates (J-gates) midway between the
A-gates. The overlap of wave functions of localized electrons in Si drops
very rapidly with distance $r$, $\exp (-2r/a_{\mathrm{B}})$, where 
$a_{\mathrm{B}}$ is the radius of localization (for P in Si, 
$a_{\mathrm{B}}\approx 2.5$~nm), therefore the interqubit distance $r$ 
between P atoms must be small (less than 20~nm) to allow overlap.

Experimental realization of the suggested model presents a number of
difficulties. We focus here on two problems: (1) placing single P donors
into the Si substrate at a precise depth underneath the barrier and (2) the
necessity to increase significantly the interqubit distance $r$ to have room
enough to arrange the metallic gates. This means that mechanisms other than
the direct overlap of the electron wave functions have to be chosen for the
coupling of adjacent nuclear spin qubits.

To solve these problems, we suggest the novel technology based on epitaxial
growth of Si and SiGe layers from isotopically engineered Si and Ge sources
followed by neutron-transmutation doping of the grown structures. We also
describe the mechanism of indirect interqubit coupling based on the
arrangement of qubits in a mesoscopic quasi-one-dimensional wire. This
mechanism allows one to control the coupling between qubits separated by
large distances $r$ (200~nm or even more), which permits the fabrication of
metallic gates by means of the modern lithography. Moreover, the suggested
mechanism of indirect coupling allows entanglement passing over non-qubit
nodes in an array of qubits. We also present the numerical estimates which
justify the feasibility of the proposed device and method of operation.

\begin{figure*}[t]
\begin{center}
\includegraphics[width=16cm]{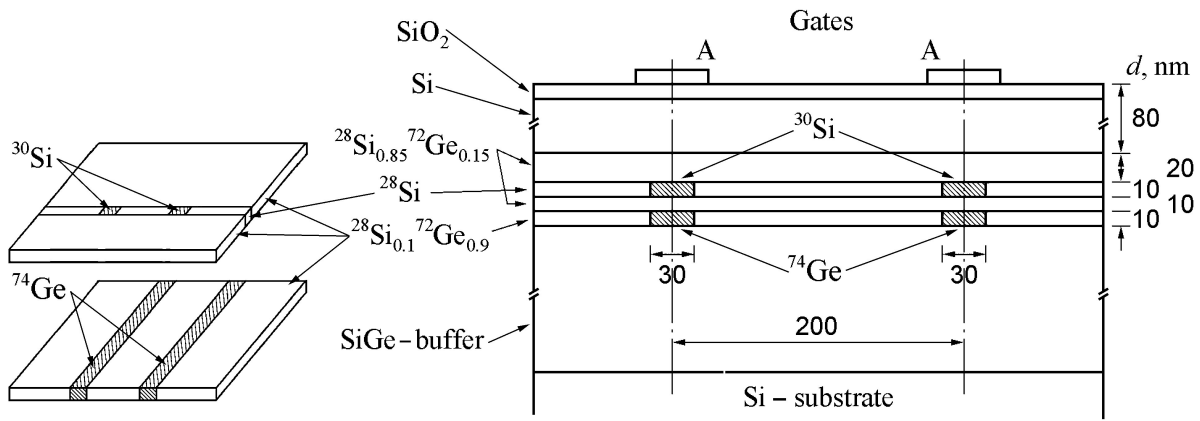}
\end{center}
\caption{Schematics of the proposed device. After NTD, ${}^{31}$P donors
appear only inside the ${}^{30}$Si-spots and underlying ${}{}^{74}$Ge-strips
will be heavily doped with ${}^{75}$As donors. All sizes are shown in nm.}
\end{figure*}

\section{Device fabrication}

\subsection{Precise placement of P atoms into Si}

Two methods had prevoiusly been suggested for the precise placing of P atoms
into Si (for a review, see Ref.~\onlinecite{Clark} and references therein).
In \cite{Brien} and \cite{Schofield}, a ''bottom-up'' method is described
for incorporation of phosphorus-bearing molecules PH$_{3}$ on a preliminary
H passivated Si (001) surface during the molecular-beam-epitaxial growth
followed by the decomposition of PH$_{3}$ at significantly increased
temperature. The alternative ''top-down'' method is based on incorporation
of dopant atoms under the surface of the grown structure using ion
implantation, followed by the annealing of radiation damage 
\cite{Yang,Park,Persaud} at increased temperatures. In these methods, the depth
distribution of the incorporated donor atoms cannot be controlled to the
necessary accuracy. For example, in the ''bottom-up'' method, incorporation
of impurities of different chemical nature is disadvantageous both for the
quality of the growing structure and for the sharpness of the vertical
distribution of impurities because of the ''floating-up'' effect in the
process of subsequent growth. In the ''top-down'' method, the distribution
of impurities in ''as-implanted'' samples is determined by the energy of
ions, and initially is not sharp. Moreover, during the annealing of
radiation damage, impurities are redistributed towards the Si/SiGe interface
which acts as a getter. As a result, the vertical distribution of the
implanted impurities is rather extended. Meanwhile, small fluctuations in
the vertical position of P atoms under the gate is very important to
minimize the A-gate voltage error rate \cite{Larionov} because otherwise,
each qubit would need its own set of applied voltages and each interacting
pair of qubits would need its unique pulse time \cite{Keyes}.

We suggest here a novel technology of the precise placing of P atoms into Si
layer. The key point is the growth of the central Si and barrier 
Si$_{0.85}$Ge$_{0.15}$ layers from different isotopes: the Si$_{0.85}$Ge$_{0.15}$
layers from isotopes $^{28}$Si and $^{72}$Ge and the central Si layer from
isotope $^{28}$Si with $^{30}$Si spots introduced by means of the
nano-lithography (Fig.~1). The formation of quasi-one-dimensional Si wires
will be achieved in a subsequent operation by the etching of Si layer
between wires and the filling of the resulting gaps by the Si$_{0.85}$Ge$%
_{0.15}$ barrier composed from isotopes $^{28}$Si and $^{72}$Ge. Because
different isotopes of Si and Ge are chemically identical, this technology
guarantees the high quality of the grown structures.

After preparation, these structures will be irradiated with a neutron flux
in a nuclear reactor followed by the fast annealing of radiation damage. The
behaviour of different isotopes is different. After capture of a slow
(thermal) neutron, a given isotope shifts to the isotope with mass number
larger by one. If the isotope thus obtained is stable, this nuclear reaction
does not entail doping. However, if the obtained isotope is unstable, it
transmutes after half-life time $\tau $ to a nucleus of another element with
atomic number larger by one in the case of $\beta ^{-}-$\ decay. This method
of doping is called NTD - neutron-transmutation doping \cite{NTD}. In the
case of Si, NTD is based on the transmutation of the isotope $^{30}$Si: 
\begin{equation*}
{}_{14}^{30}\mathrm{Si}+{}_{0}^{1}n={}_{14}^{31}\mathrm{Si}\rightarrow \beta
^{-}(\tau =2.62h)\rightarrow {}_{15}^{31}\mathrm{P}.
\end{equation*}

In the isotopically engineered structure, $^{31}$P donor atoms will be
produced only within $^{30}$Si spots, because the isotopes $^{28}$Si and $%
^{72}$Ge shift to the stable isotopes $^{29}$Si and $^{73}$Ge, respectively.
Therefore, in the suggested method, the processes of the structure growth
and doping are completely separated. The idea to fabricate a basic element
for a nuclear spin quantum computer using the isotope engineering of Si and
SiGe nanostructures was proposed earlier in \cite{Shlimak1,Ladd}. The
precise placing of P atoms into a Si matrix by means of the NTD method was
suggested in \cite{Shlimak2}.

Let us estimate the feasibility of the proposed method. We consider a 
$^{30}$Si spot of area $S=30\times 30$~nm $\approx 10^{-11}$~cm$^{2}$, thickness 
$d=$10~nm with a distance of 200~nm between spots (Fig.~1). The ''buried''
distance to the structure surface is halved, 100 nm, which is needed to
protect against cross-links and to ensure the influence of the A-gate
voltage on the corresponding underlying qubit only. In the proposed method
of incorporation of P into Si, the vertical accuracy of the location of P
donors is determined by the thickness of the $^{30}$Si spot (10~nm) with
respect to the distance to the A-gate (100~nm). Indeed, the irradiation of
samples by thermal neutrons occurs at room temperatures where the introduced
P atoms are immobile. The irradiation is followed by the annealing of
radiation damage at higher temperatures (700$^{\circ }$C). During the short
annealing time (1 hour), impurities cannot diffuse far from the
transmutation place; the diffusion length does not exceed 1--2~nm. As a
result, the proposed method will provide an almost equal burying depth of P
atoms with a controlled accuracy (about 10\thinspace \% in our example). The
time needed for the irradiation of the grown structures in a nuclear reactor
is estimated as follows. The number of transmutation events is 
$\widetilde{N}(^{31}\mathrm{P})=\widetilde{N}(^{30}\mathrm{Si})\sigma^{30}\Phi$, 
where $\widetilde{N}$ is the number of atoms, $\sigma $ is the cross-section of the
thermal neutron capture for given isotope 
($\sigma ^{30}\approx 0.11\cdot 10^{-24}$~cm$^{2}$), $\Phi =\varphi t$ is 
the integral neutron flux ($\varphi $ is the intensity of the thermal 
neutron flux and $t$ is the time
of irradiation). In a spot with volume $V=30\times 30\times 10$~nm 
$\approx 10^{-17}$~cm$^{3}$, there are 
$\widetilde{N}=N(\mathrm{Si})V\approx 5\cdot 10^{5}$~Si atoms 
($N(\mathrm{Si})=5\cdot 10^{22}$~cm$^{-3}$). If the
enrichment of Si with isotope $^{30}$Si is close to 100\thinspace \%, 
$\widetilde{N}(^{30}\mathrm{Si})\approx 5\cdot 10^{5}$. To achieve 
$\widetilde{N}(^{31}\mathrm{P})=1$, the integral irradiation dose $\Phi $ has
to equal to $2\cdot 10^{19}$~neutron/cm$^{2}$. In some research nuclear
reactors, $\varphi \leqslant 10^{14}$ cm$^{-2}$ s$^{-1}$, which corresponds
to the reasonable irradiation time $t\approx 2\cdot 10^{5}$~s 
$\approx 56$~hrs.

\subsection{Qubit certification}

An unavoidable peculiarity of the NTD is the casual character of the neutron
capture. As a consequence, after NTD, some of $^{30}$Si spots will contain
no donor atom (``0-spot'') and cannot serve therefore as qubits, while some
will contain more than one donor atom. The probability $P_{m}$ to find
``0-spot'', ``1-spot'', ``2-spot'', and so on $(m=0,1,2,...)$, is described
by the Poisson distribution: 
\begin{equation*}
P_{m}=C_{D}^{m}p^{m}(1-p)^{D-m}
\end{equation*}%
where $C_{D}^{m}$ is the binomial coefficient, 
$p=N(^{30}\mathrm{Si})\sigma ^{30}d$ is the probability for the neutron 
to be captured in a layer
of thickness $d$, \ $D=\Phi S$ is the dimensionless dose of irradiation.

\begin{figure}[tb]
\begin{center}
\includegraphics[width=8cm]{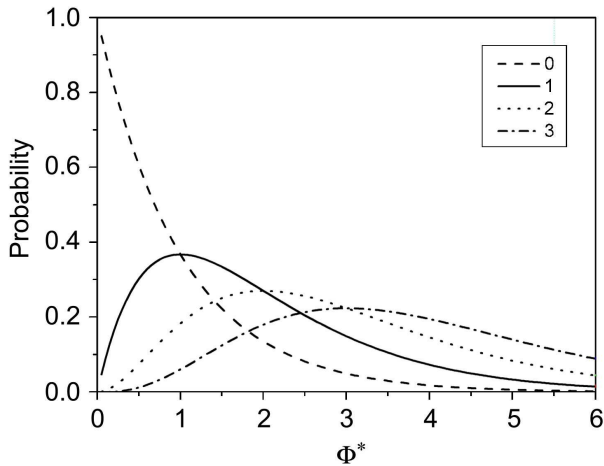}
\end{center}
\caption{Probability to find spots with 0, 1, 2 and 3 impurity atoms,
introduced by NTD, as a function of the dimensionless irradiation dose $\Phi
^{\ast }$.}
\end{figure}

These probabilities are plotted in Fig. 2 as a function of the dimensionless
parameter $\Phi ^{\ast }=pD.$ The best candidates for qubits are
``1-spots'', while spots having more than one donor could also be considered
as qubits after corresponding theoretical investigation. Only spots without
donors (``0-spots'') are obviously non-qubits. In the NTD method, the
maximal portion of ``1-spots'' is 37\thinspace \% at $\Phi ^{\ast }=1$. In
this case, the portion of non-qubit is also about 37\thinspace \%. If
``2-spots'' are also considered as possible qubits, the optimal integral
dose will correspond to $\Phi ^{\ast }=2.$ In this case, the fraction of
non-qubits decreases to $\approx $ 14\thinspace \%.

To determine the number of donors in each spot, we propose to use narrow
source-drain (SD) channels fabricated beneath each donor position (Fig.~1).
If the given spot contains one donor, it will form (together with the
underlying SD channel and overlying A-gate) a flash-memory field-effect
transistor (FET) with the qubit donor acting as a ''floating gate''. The
donor electron is separated from the SD channel and A-gate by the Si$_{0.85}$Ge$_{0.15}$ 
barriers of about 10--20~nm width and 100 meV height \cite{Kane1}
which are used for the electron confinement. However, a pulse of relatively
strong voltage applied between the A-gate and SD channel will tilt the
barriers leading to electron transfer and donor ionization. It was shown in
Ref.~\onlinecite{Smit} that if the dopant-gate separation distance is more
than 8$a_{\mathrm{B}}\approx 24$~nm for P in Si (in our case this condition
is satisfied because the distance to the gate is about 100 nm), the transfer
of electron from donor takes place abruptly at a threshold voltage. An
electric field of the positively-charged donor ion located only 10 nm from
the narrow SD channel will shift the FET cutoff voltage at the value of
about 10 mV, which is easy to observe \cite{Streetman}. If there are two or
more donors in the given spot, the cutoff shift will be even larger. If
there is no donor underneath the given gate, the shift will not be observed.

\subsection{Read-out operation}

We believe that the proposed SD channel can also be used for the read-out
operation, i.e. for the detection of a single nuclear spin state. The direct
control of a nuclear spin state via nuclear magnetic resonance (NMR)
measurements is a difficult problem. In Ref.~\onlinecite{Machida}, coherent
control of the local nuclear spin was demostrated, based on pulsed NMR in a
quantum Hall device. In Ref.~\onlinecite{Yusa}, a self-contained
semiconductor device is described that can control nuclear spins in a
nano-scale region. Measurements of the electron spin state are much easier
taking into account the possibility of a spin-to-charge conversion for
electrons. In accordance with the Kane model \cite{Kane98}, the state of the
nuclear spin $^{31}$P is mediated by the spin of donor electron via the
hyperfine interaction. Therefore, the task is to determine the spin
orientation of the corresponding donor electron. The suggested method 
\cite{Vrijen} is based on the fact that at low temperatures, a donor atom can
capture the second electron with small ionization energy, about 1~meV, which
results in the appearance of a negatively charged donor ($D^{-}$-center).
However, this process is possible only when the spin orientation of the
second electron is opposite to that of the first electron. The appearance of
the charged donor in the vicinity of the narrow SD channel will affect the
current \cite{Kurten,Xiao} and can therefore be detected. As a result, one
can determine the spin orientation of two neighboring donor electrons if one
applies a potential difference between the corresponding A-gates which will
cause the electron to jump from one donor to another . If we choose the spin
orientation of the given donor as a reference, one can determine the spin
state of the neighboring qubits on the right and left sides.

\subsection{Source-drain channels fabrication}

The proposed SD channels have a twofold purpose: the determination of the
number of donors within each spot (qubit certification) and the
determination of the spin state of the donor electron (read-out operation).
We suggest a method for the fabrication and the proper alignment of the SD
channels with respect to the position of $^{31}$P qubits. This method is
also based on isotope engineering of a Si$_{1-x}$Ge$_{x}$ layer followed by
NTD. We propose to make the underlying Si$_{1-x}$Ge$_{x}$ layer from a
composition close to pure Ge (say, Si$_{0.1}$Ge$_{0.9}$) using isotopes 
$^{28}$Si and $^{72}$Ge, followed by the fabrication of 30 nm-width strips
where $^{72}$Ge is replaced by $^{74}$Ge (Fig.~1). After NTD, these 
$^{74}$Ge-strips will be doped by As donors through the following nuclear reaction: 
\begin{equation*}
{}_{32}^{74}\mathrm{Ge}+{}_{0}^{1}n={}_{32}^{75}\mathrm{Ge}\rightarrow \beta
^{-}(\tau =82\min )\rightarrow {}_{33}^{75}\mathrm{As}
\end{equation*}

Irradiation of the structure with the thermal neutron integral dose 
$\varphi t=2\cdot 10^{19}$~cm$^{-2}$ ($\Phi ^{\ast }=1$) needed for introducing on
average one P donor in each $^{30}$Si spot, will also lead to doping of 
$^{74}$Ge-strips with As donors to a high level 
($N_{\text{As}}\approx 4.5\cdot 10^{17}$~cm$^{-3})$ because of the relatively large 
$\sigma^{74}=0.5\cdot 10^{-24}$~cm$^{2}$. This concentration of As exceeds the
critical value of the metal--insulator transition for Ge:As \cite{Shlimak}.
Therefore, NTD-introduced narrow channels will have a metallic-like
conductivity and remain conductive down to $T\rightarrow 0$. This is
important because nano-FET will operate at low temperatures when donor
electrons in $^{31}$P-qubits are localized on their donors. For the
suggested geometry of the SD channel, with the thickness of the 
Si$_{0.1}$Ge$_{0.9}$ layer of about 10~nm, width of the $^{74}$Ge-strips of 30~nm and the
length about 1~$\mu $m, the channel resistance is about 1~M$\Omega $, which
is suitable.

In the proposed method of device fabrication, the proper alignment of the 
$^{74}$Ge-strips with respect to the overlying $^{31}$P qubits is provided by
the high accuracy of the electron beam-assisted patterning of trenches in 
$^{72}$Ge$^{28}$Si layer with the overlying holes in $^{28}$Si layer. Taking
into account that the size of all components is not less than 30--50~nm, one
can conclude that the alignment could be realized by the recent progress in
SEM- and AFM-assisted lithography.

\begin{figure}[tb]
\begin{center}
\includegraphics[width=6cm]{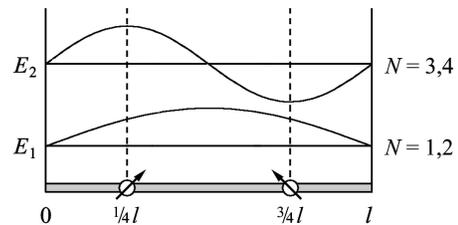}
\end{center}
\caption{Spatial distribution of electron wave functions for $N$ electrons
in a quantum wire of the length $l$.}
\end{figure}

\section{Two-qubit operation}

In this section, we suggest a new mechanism of entanglement for distant
qubits and discuss, first, the principles of two-qubit operation. It has
been shown \cite{DiVincenzo-2} that two-bit gates applied to a pair of
electron or nuclear spins are universal for the verification of all
principles of quantum computation.

Because direct overlap of wavefunctions for electrons localized on P donors
is negligible for distant pairs, we propose another principle of coupling
based on the placement of qubits at fixed positions in a
quasi-one-dimensional Si nanowire and using the indirect interaction of 
$^{31}$P nuclear spins with spins of electrons localized in the nanowire
which we will call hereafter as ''1D-electrons''. This interaction depends
on the amplitude of the wavefunction of the ''1D-electron'' estimated at the
position of the given donor nucleus $\Psi _{n}(r_{i})$ and can be controlled
by the change in the number of ''1D-electrons'' $N$ in the wire$.$

At $N=0,$ the interqubit coupling is totally suppressed, each $^{31}$P
nuclear spin interact only with its own donor electron. This situation is
analoguous to that one suggested in the Kane proposal \cite{Kane1} and
therefore all single-qubit operations and estimates of the decoherence time
are valid also in our case. The difference consists in the method of
coupling when a controlled number $N$ of ''1D-electrons''is injected into
the nanowire. In this case, nuclear spin-qubits will also interact with the
spins of ''1D-electrons''. To estimate the intensity of this interaction we
need to calculate $\Psi _{n}(r_{i}).$ In the below calculation we believe
that the donor potential does not influence the distribution function $\Psi
_{n}(r)$ of ''1D-electrons'' because it is screened by the donor electron on
the relatively short distance of order of $a_{\mathrm{B}}\approx 2.5$~nm and 
$a_{\mathrm{B}}$ is small comared with the wire length $l$.

Let the interqubit distance be $r=200$~nm, one order of magnitude larger
than in the Kane proposal \cite{Kane98}. To realize the coupling between
these distant qubits, we suggest fabricating a Si nanowire of length 
$l=400$~nm and place P donors at distances $r_{1}=(1/4)l$, and $r_{2}=(3/4)l$ 
(see Fig.~3). For $N=1$ and $N=3,$ the functions $\Psi _{n}(r)$ belong to the
energy levels $E_{n}$ ($n=1,2$) because each level contains two electrons
with opposite spin. The highest amplitude of $\Psi _{n}(r)$ evaluated at the
positions of the nuclear spin qubits $r_{1}$ and $r_{2}$ is realized at $N=3$
$(n=2)$. In this case, the interqubit coupling is maximal.

To calculate the coefficient of the hyperfine interaction between nuclear
and electron spins, we consider following \cite{PVW03}, a system consisting
of electrons confined by a potential $V\left(\vec{r}\right)$ and two
nuclear spins. We suppose that the nuclear spins are located far enough from
each other so that the direct nuclear spin interaction is negligible. The
contact hyperfine interaction between electrons and nuclear spins leads to
an indirect nuclear spin interaction. Let the quantum wire have finite
length $l$ in the $x$-direction with the two nuclear spins located at 
$\vec{r}_{1}$ and $\vec{r}_{2}$ in a magnetic field $\mathbf{H}$ directed in 
the $z$ direction. The Hamiltonian of the system is given by 
\begin{multline}
H=H_{0}+H_{1}=\frac{1}{2m_{e}}\left( \mathbf{p+}\frac{e}{c}\mathbf{A}\right)
^{2}+V\left( \vec{r}\right) \\
-2g\mu _{B}\mathbf{\sigma H}+\frac{8\pi }{3}\mu _{B}\gamma _{n}\hbar 
\mathbf{I}_{1}\mathbf{\sigma }\delta \left( \vec{r}-\vec{r}_{1}\right) \\
+\frac{8\pi }{3}\mu _{B}\gamma _{n}\hbar \mathbf{I}_{2}\mathbf{\sigma}
\delta \left( \vec{r}-\vec{r}_{2}\right) \label{Hamit}
\end{multline}
where $H_{0}$ is the Hamiltonian of the electron in the mesoscopic structure
in the magnetic field, $H_{1}=H_{1}^{\left( 1\right) }+H_{1}^{(2)}$ is the
contact hyperfine interaction, $m_{e}$ is the electron mass, $\mathbf{A}$ is
the magnetic-field potential, $\mu _{B}$ is the Bohr magneton, $\gamma _{n}$
is the nuclear gyromagnetic ratio, $\mathbf{I}_{1,2}$ and $\mathbf{\sigma }$
are nuclear and electron spins, and $\vec{r}_{1,2}$ is radius vector of the
nucleus.

The effective nuclear spin interaction energy calculated by second-order
perturbation theory is \cite{Abragam}: 
\begin{equation}
E=\sum_{E_{i},E_{f}}\frac{\left\langle \Psi _{i}\left| H_{1}^{(1)}\right|
\Psi _{f}\right\rangle \left\langle \Psi _{f}\left| H_{1}^{(2)}\right| \Psi
_{i}\right\rangle }{E_{f}-E_{i}}f_{i}\left( 1-f_{f}\right) +c.c.
\label{eff_en}
\end{equation}%
Here, $\Psi _{i}$ and $E_{i}$ are the eigenfunctions and eigenvalues of $%
H_{0},$ and $f_{i,f}$ is the electron distribution function in the initial
and final states. We will use expression (\ref{eff_en}) to find the
effective interaction between nuclear spins.

We suppose that the transverse dimensions of the quantum wire are much
smaller than its length and the cyclotron orbit of the electron. The
confining potential is $V(x,y,z)=V\left(x\right)-V_{0}\delta (y)\delta (z)$,
where $V(x)=0$ if $0\leqslant x\leqslant l$, and $V(x)=\infty $ otherwise.

The wave function should satisfy the following boundary condition: 
$\Psi_{n}\left(0\right)=\Psi_{n}\left(l\right)=0$. The solution has the form 
\begin{multline}
\Psi _{m,s=\pm }=\sqrt{\frac{4}{\pi l\delta y\delta z}}\left( \left( 
\begin{array}{c}
1 \\ 
0
\end{array}
\right) ,\left( 
\begin{array}{c}
0 \\ 
1
\end{array}
\right) \right) \\
\times\sin \left( \frac{n\pi x}{l}\right)
 \exp \left( -\frac{y^{2}}{\delta y^{2}}\right) \exp \left( 
-\frac{z^{2}}{\delta z^{2}}\right) ,
\end{multline}
\begin{equation}
E_{n,s=\pm }=\frac{\hbar ^{2}\pi ^{2}}{2m_{e}l^{2}}n^{2}\mp g\mu _{B}H,
\label{enlevel}
\end{equation}%
where $\delta y$ and $\delta z$ are the transverse dimensions of the
electron wave function.

Let us consider the problem at $T=0$. In this case the electron distribution
function $f$ is 1 for the filled states and 0 for the empty states.
Inserting this wave function into Eq. (\ref{eff_en}) and assuming that the
Zeeman splitting energy in (\ref{enlevel}) is much less than the energy gap
between levels with different $n$, one obtains the following expression for
the nuclear spin interaction constant $A$ \cite{PVW03}: 
\begin{equation}
A=\left( \frac{32\mu _{B}\gamma _{n}\hslash }{3l\delta y\delta z}\right) ^{2}%
\frac{\sin ^{2}\left[ \frac{N+1}{2}\pi (r_{1}/l)\right] \sin ^{2}\left[ 
\frac{N+1}{2}\pi (r_{2}/l)\right] }{g\mu _{B}H}  \label{A}
\end{equation}

It is seen, that at $N=3$, $A$ is maximal for both qubit positions 
$r_{1}=(1/4)l$ and $r_{2}=(3/4)l$. Let us estimate the error caused by
unavoidable fluctuations in the positions of nuclear spins in the wire. In
our device, the size of $^{30}$Si-spot is 30 nm. Therefore, one can expect
that the position of NTD-introduced P donor will fluctuate around the
central point within $\pm 15$~nm, which is about 1/40 of the total wire
length (600 nm), $\Delta r/l\leq 0.025$. In our model, the coupling is
realized in the case when the wave function of ''1D-electron'' $\Psi (r)$
has the maximal value at places of the qubit location, where the space
derivative $d\Psi /dx$ is close to zero. This makes the coupling insensitive
to the form of distribution function and to small fluctuations in the qubit
positions. 

Thus, the above consideration shows that the indirect coupling is maximal at 
$N=3,$ while at $N=0,$ the coupling is totally suppressed. This means that
in our model, the entanglement between two distant qubit can be effectively
controlled by the proper variation of $N$. 

\section{Scalability}

Scalability is the one of the most important requirements of the quantum
computer proposals \cite{DiVincenzo}. We suggest below the schematics of the
device architecture (Fig.~4) which satisfy the scalability requirements. It
is worth mentioning that the above method of coupling opens a way to avoid
the problem connected with the break in the one-dimensional array of qubits.
This problem is inevitable in all proposed technologies. In the method of
coupling based on the direct overlap of electron wave functions \cite{Kane98,Vrijen}, 
this requirement is crucial because any break in the
one-dimensional array of qubits stops the entanglement along the array and
make quantum computation impossible. In our model, entanglement can exist
even in the unlikely case of two or more breaks in the qubit array one after
another, because indirect coupling can passing over wrong sites by the
proper choice of the nanowire length $l$ and the number of electrons $N$ in
the wire.

Figure 4 shows the schematics of the device architecture which allows one to
vary $l$ and $N$. The device consists of a $^{28}$Si nanowire with an array
of $^{30}$Si spots. Each spot is supplied by the overlying A-gate, the
underlying SD-shannel and the lateral N-gate. After NTD, P donors will
appear in most of the spots (which transforms these spots into qubits) and
not appear in other spots (non-qubits). This situation is shown
schematically in Fig. 4 where one assume that the spots 3 and 4 are
non-qubits (''0-spots'') and one need to provide coupling between qubits $2$
and $5$. For this purpose, it is necessary to connect the gates N$_{2}$,
N$_{3}$, N$_{4}$ and N$_{5}$. The negative voltage applied between other
N-gates and the wire contact $L$ will lead to pressing-out ''1D-electrons''
from all corresponding areas and formation of the nanowire with $l=800$~nm
between the sites 2 and 5 only (shown in grey in Fig. 4). The coupling
between qubits $2$ and $5$ will be realized via injection in the wire of the
necessary number of electrons $N$, using the positive voltage applied to the
gates N$_{2}$--N$_{5}$. In this particular example, the maximal coupling
will be realized at $N=7$, while at $N=0$, the coupling will be totaly
suppresed.

\begin{figure}[tb]
\begin{center}
\includegraphics[width=8cm]{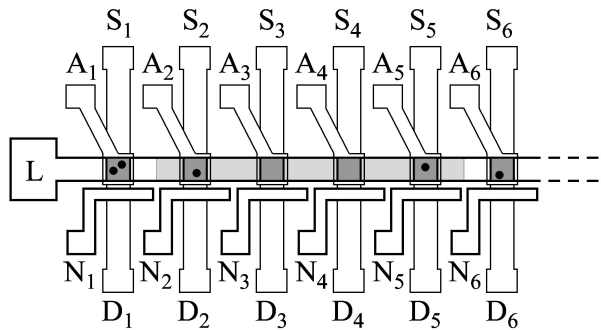}
\end{center}
\caption{{}Schematics of a $^{28}$Si nanowire $L$ with an array of $^{30}$Si
spots (qubits and non-qubits after NTD). Each spot is supplied by overlying
A-gate, underlying SD-channel and lateral N-gate. This device architecture
allows to realize an indirect coupling between any distant qubits (see
text). }
\end{figure}

\section{\protect\bigskip Summary}

A new method of a quantum information processing is suggested based on the
precise placing of $^{31}$P nuclear spin qubits in a quasi-one-dimensional $%
^{28}$Si nanowire. The fabrication method includes the isotope engineering
of Si and SiGe layers followed by the neutron-transmutation doping of the
obtained structures. The interqubit coupling is based on the indirect
interaction of $^{31}$P nuclear spin with the spin of electron localized in
the nanowire. The advantage of the proposed method of operation consists in
its ability to control the coupling between distant qubits and even between
qubits separated by non-qubits nodes in a one-dimensional array. The last
feature allows one to develop the basic unit and the scalable architecture
of a nuclear spin-based quantum computer. Numerical estimates show the
feasibility of the proposed methods.

\section{Acknowledgements}

We are thankful to P. Wyder and Yu.V. Pershin for fruitful discussions. I.S
thanks the Erick and Sheila Samson Chair of Semiconductor Technology for
financial support. I.V. acknowledges the support of the Brussels program
EuroMagNET RII3-CT-2004-506.

\end{document}